\documentclass{moriond}

\usepackage{amsmath}

\def\be{\begin{equation}}
\def\ee{\end{equation}}
\def\bea{\begin{eqnarray}}
\def\eea{\end{eqnarray}}

\begin{document}
\vspace*{4cm}
\title{How to fit PDFs in the presence of new physics?}

\author{ Elie Hammou }

\address{DAMTP, University of Cambridge,\\ Wilberforce Road, Cambridge, CB3 0WA, UK}

\maketitle\abstracts{
The interpretation of LHC data, and the assessment of possible hints of new physics, require the precise
knowledge of the proton structure in terms of parton distribution functions (PDFs). I present a
methodology designed to determine whether and how global PDF fits might inadvertently 'fit away'
signs of new physics in the high-energy tails of the distributions. I showcase a scenario for the High-Luminosity
LHC, in which the PDFs may completely absorb such signs of new physics, thus biasing theoretical predictions
and interpretations. I discuss strategies to single out the effects in this scenario and disentangle the
inconsistencies that stem from them. This study brings to light the synergy between the high luminosity
programme at the LHC and present and future low-energy non-LHC measurements of large-sea quark distributions. I also present the potential of fitting simultaneously the PDFs and the new physics signals.}

\section{Background and motivation}
\label{sec_1:motivation}


The Parton Distribution Functions (PDFs) describe the proton in terms of its partonic content, the quarks and gluons. They depend on two parameters: the energy scale $Q$ and the Bjorken $x$. The $Q$ dependence is well described by the DGLAP equation while the $x$ dependence is governed by non-pertubative QCD and thus needs to be fitted from data. The usual practice for PDF fitting is to assume that the underlying physical law of Nature is the Standard Model (SM) and this is what is used to compute theory predictions in the fit. However, if this were not true, and the true underlying law included some New Physics (NP), would we be at risk of absorbing such signals in the PDFs, because of our assumptions biased toward the SM?

\begin{figure}[ht!]
    \centering
    \includegraphics[width=0.49\textwidth]{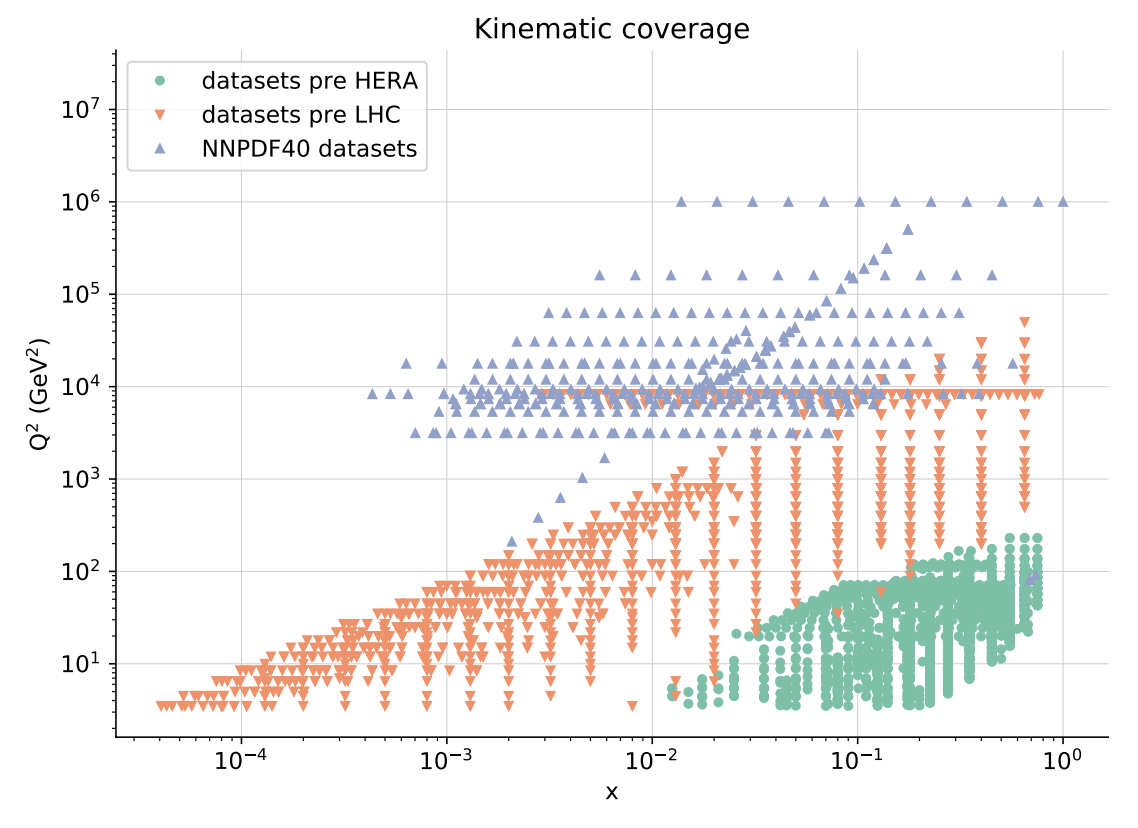}
    \includegraphics[width=0.49\textwidth]{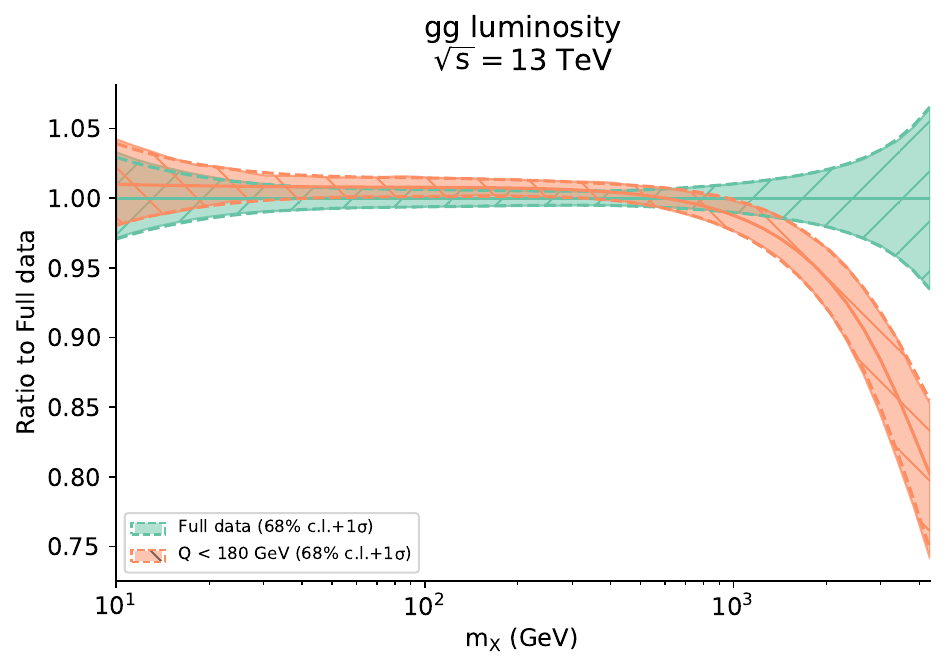}
    \caption{Left: Kinematic coverage of the full NNPDF4.0 dataset. Right: Comparison of the gluon-gluon PDF luminosities fitted on the full dataset (green) and with an energy cutoff at $Q<180$ GeV (orange).}
    \label{fig:sec1_motivation}
\end{figure}

This concern is motivated by an observation presented in Fig.~\ref{fig:sec1_motivation}. I show there two PDF fits performed using the NNPDF framework~\cite{NNPDF:2021njg}. One of those fit is run of the full dataset (green band), roughly a third of which comes from the LHC. The second one has only been trained on observables with $Q<180$ GeV (orange band), a large part of the LHC data has been excluded by this energy cutoff. We observe a discrepancy between the two $gg$ luminosities computed from the PDFs, which appears to be growing with energy. In principle, there could be several technical reasons explaining it, such as some missing higher order uncertainties or some experimental incompatibilities in the data. Another possibility is that some high-energy measurements are somewhat impacted in the high-energy tails of the distributions by some potential unknown heavy NP. The NP induced shifts in the data could then be inadvertently absorbed within the PDFs parametrisation, causing the observed shift.

In this study, I present a toy model in which we have manually injected NP signals in the high-energy data used for PDF fits and then assessed whether it could be fitted away and essentially "contaminate" the PDFs.

\section{PDF contamination from new physics signals}
\label{sec_2:contamination}

\subsection{Methodology}
\label{sec_2_1:methodology}

This risk assessment exercise of PDF contamination from heavy NP signals is described in details in a previous publication~\cite{Hammou:2023heg}. In a nutshell, it is a three-step exercise: first, one chooses a Beyond the SM (BSM) scenario susceptible to impact the data used in the PDF fit. Second, one generates some pseudodata contaminated by the BSM physics. Finally, one runs a PDF fit on this contaminated pseudodata and compares the results of the \textit{contaminated} fit to the ones of a \textit{baseline} fit which has not seen the fake BSM data. Note that this exercise is streamlined in our publicly available software {\tt SIMUnet}~\cite{Iranipour:2022iak} \cite{Costantini:2024xae}. 

To determine whether the PDFs have been contaminated by the NP, two criteria need to be met. The contaminated and baseline fits must be somewhat incompatible {\em and} the quality of the fit must not deteriorate in the presence of the BSM pseudodata. Indeed, if it did, it would be possible to identify and exclude the contaminated data from the fit.

In this study, we have chosen a $W'$ model as our BSM scenario, featuring a heavy field transforming as a triplet under $SU(2)_L$. For a mass $M_{W'}$ large enough, it can be matched to a SM Effective Field Theory (SMEFT) Lagrangian whose corrections are dominated by the four-fermion operators:

\begin{equation}
	\mathcal{L}^{W'}_{\text{SMEFT}} = \mathcal{L}_{\text{SM}} -\frac{g_{W'}^{2}}{2 M^2_{W'}} J^{a, \mu}_L J^a_{L, \mu}, \qquad J^{a, \mu}_L = \sum_{\substack{f_L}} \bar{f_L} T^a \gamma^{\mu} f_L \, .
\end{equation}

Using this model, we have generated High-Luminosity LHC (HL-LHC)  projections for Drell-Yan (DY) processes, which can be used in a PDF fit.

\subsection{Contamination and consequences}
\label{sec_2_2:missing_NP}

\begin{figure}[ht!]
    \centering
    \includegraphics[width=0.4\textwidth]{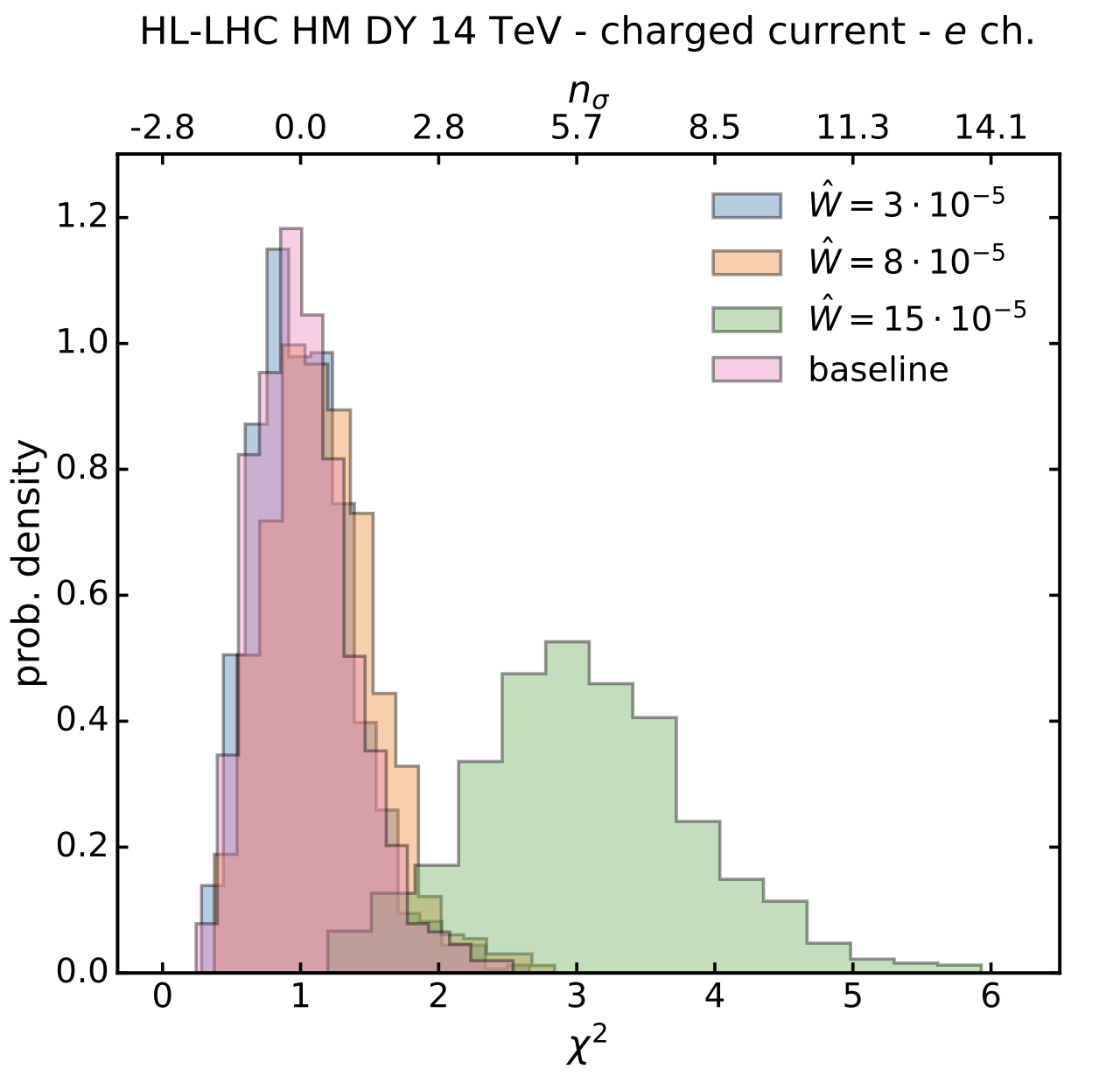}
    \includegraphics[width=0.49\textwidth]{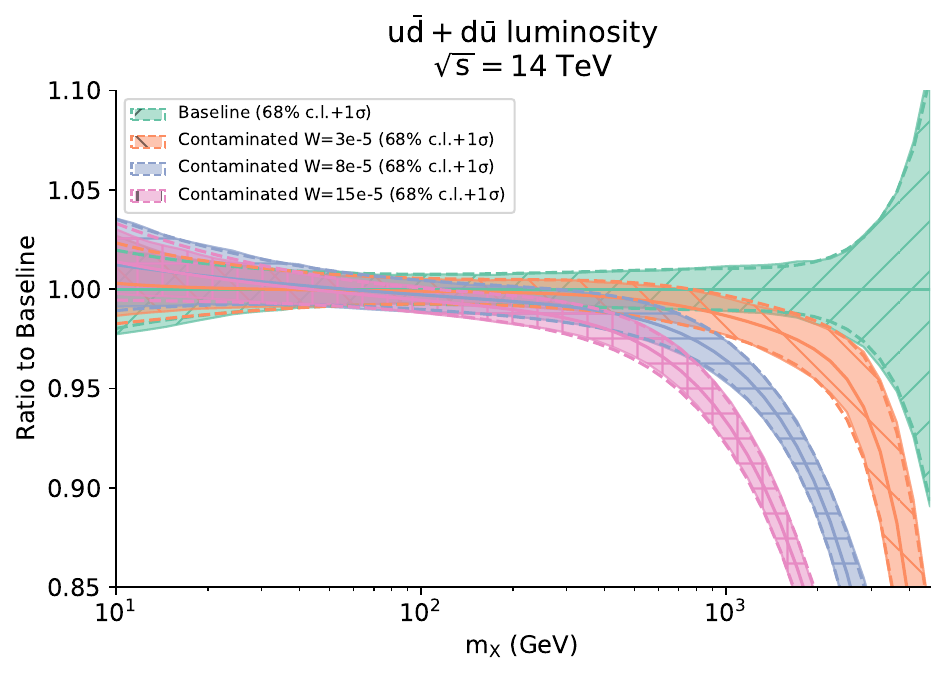}
    \caption{Left: Distribution of the $\chi^2$ values across various level-1 instances of the pseudodata for a contaminated HL-LHC dataset, measuring the quality of the PDF fit. Right: Comparison of the $u\bar{d}+d\bar{u}$ PDF luminosities of the baseline and contaminated fits.}
    \label{fig:PDF_cont}
\end{figure}

We can measure the "degree of contamination" of the data using an oblique parameter $\hat{W} \propto \frac{g_{W'}^2}{M_{W'}^2}$. The larger $\hat{W}$, the bigger the impact of the NP on the data, with $\hat{W}=0$ corresponding to the SM. We show the results of the PDF fits in the presence of different degrees of contamination in Fig.~\ref{fig:PDF_cont}. We are looking for the highest degree of contamination that does not deteriorate the quality of the fit. It is found for $\hat{W} = 8 \cdot 10^{-5}$ which corresponds to $M_{W'} = 13.8$ TeV assuming $g_{W'} = 1$. We see that for this parameter, the contaminated $u\bar{d}+d\bar{u}$ luminosity is incompatible with the baseline one, and that the shifts grows with energy similarly to what we observed in Fig.~\ref{fig:sec1_motivation}. This means that {\em the PDFs have been contaminated by the NP} in this case.

The first obvious negative consequence of this is that we are at risk of missing NP signals which are fitted away in the PDFs. Indeed, the non-deterioration of the fit quality implies that the PDFs have been warped in a way that compensates the NP shifts, making the SM theory predictions compatible with the BSM data. Furthermore, a second equally worrying consequence emerges. This warping of the PDFs risks creating fictitious deviations in other sectors which should not be impacted by the $W'$. We observed this in different diboson production processes~\cite{Hammou:2023heg}. If observed in real life, such a fake discrepancy would risk sending someone on a wild goose chase.

We have been able to identify two possible solutions to the PDFs contamination study we observed in our toy model.

\section{Possible solutions}
\label{sec_3:solutions}

\subsection{Synergy of high- and low-energy observables}
\label{sec_3_1:low-E}

\begin{figure}[ht!]
    \centering
    \includegraphics[width=0.4\textwidth]{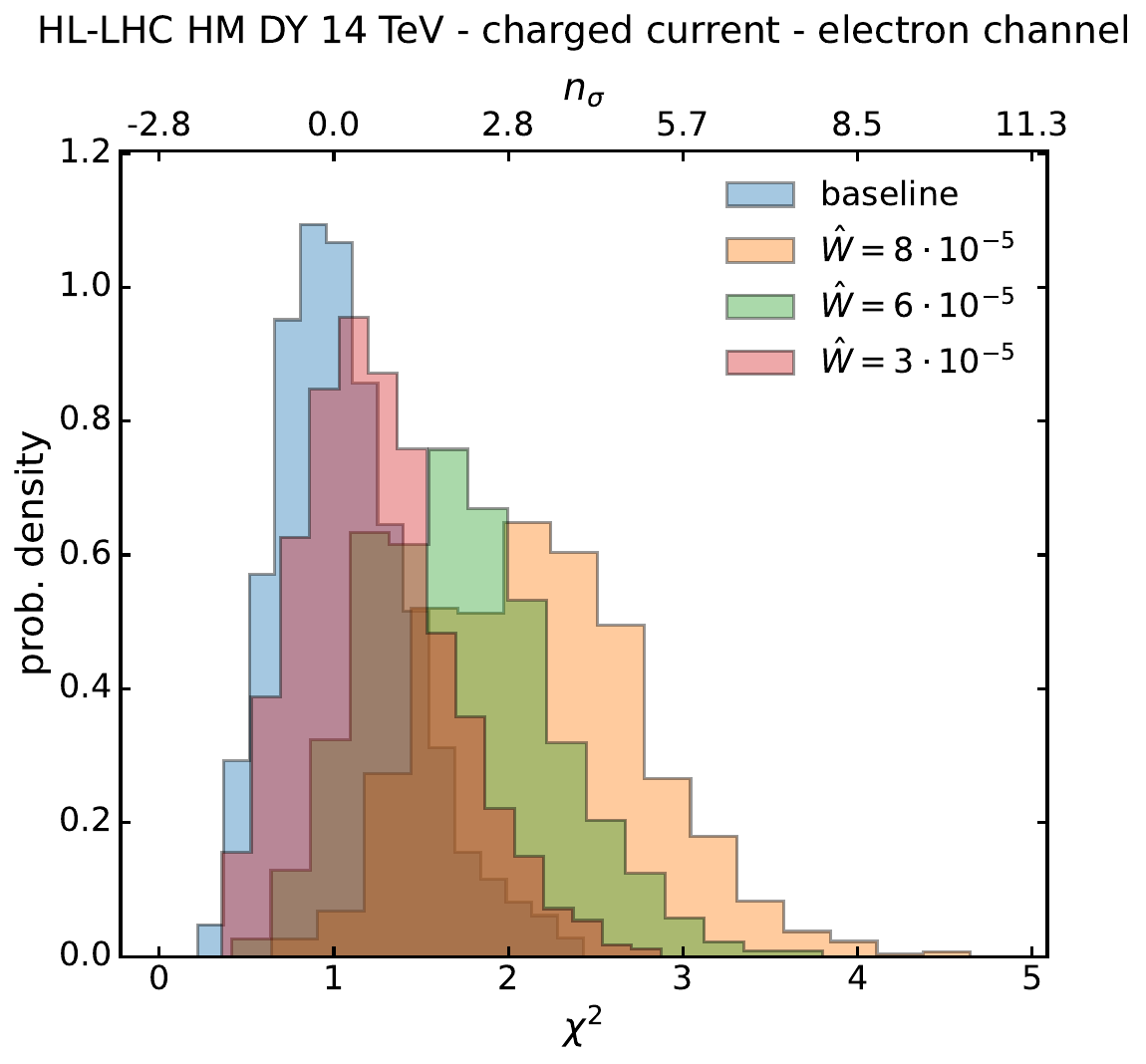}
    \includegraphics[width=0.49\textwidth]{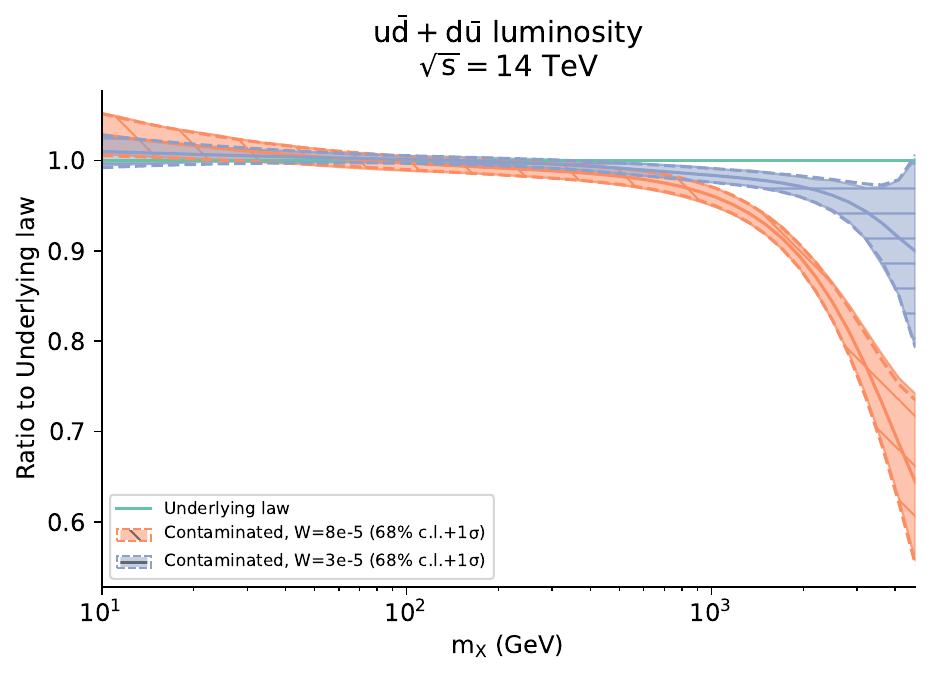}
    \caption{Left: Quality of the fit with FPF projection data. Right: Comparison of the $u\bar{d}+d\bar{u}$ luminosity fitted in the presence of NP with (blue) and without (orange) FPF projection data.}
    \label{fig:low_E}
\end{figure}

The first one is to strip the PDFs from the excessive degrees of freedom which allow them to absorb the NP signals by constraining the large-x region more tightly with low-energy data, which would be safe from heavy NP contamination. In Fig.~\ref{fig:low_E}, we see the impact of the inclusion of projection data of the Forward Physics Facility (FPF)~\cite{Anchordoqui:2021ghd} \cite{Cruz-Martinez:2023sdv}. The main result is that our previous maximally contaminating scenario is now flagged by a visible deterioration of the fit quality. The new threshold is reached at $M_{W'}=22.5$ TeV, meaning that a $W'$ is only at risk of being fitted away in the PDFs for masses greater than this. While it does not solve completely the issue, it is a significant improvement. Furthermore, the worst case scenario presented on the right plot of Fig.~\ref{fig:low_E}, the $u\bar{d}+d\bar{u}$ luminosity fitted with FPF projection data (blue) is much more compatible with the underlying law than the one fitted without (orange).

More results will be presented in a forthcoming study~\cite{Hammou:2024xxx}, including the projected impact of the Electron-Ion Collider (EIC).

\subsection{Simultaneous fits of PDFs and SMEFT operators}
\label{sec_3_2:simu_fit}

\begin{figure}[ht!]
    \centering
    \includegraphics[width=0.49\textwidth]{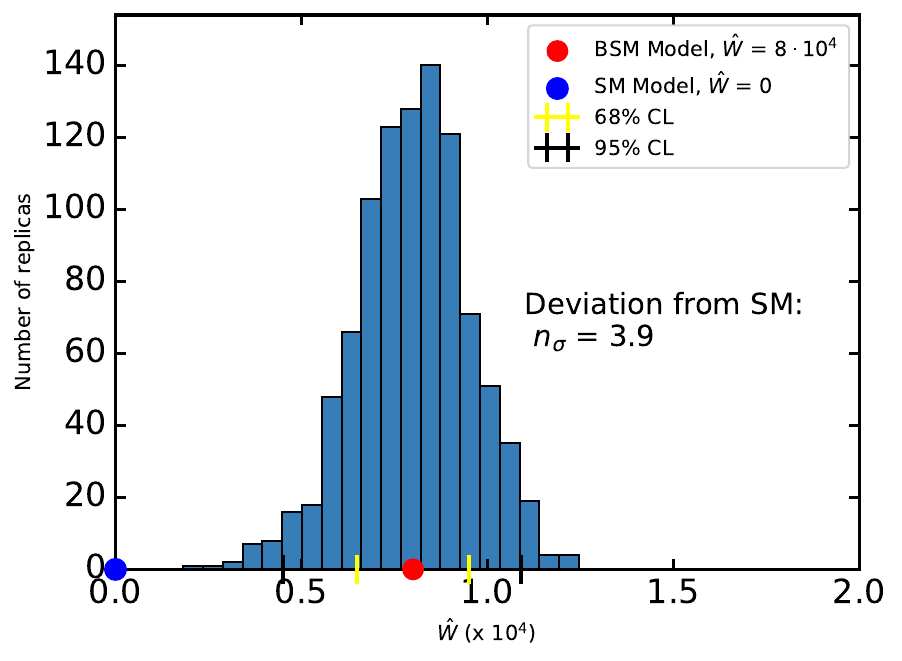}
    \includegraphics[width=0.49\textwidth]{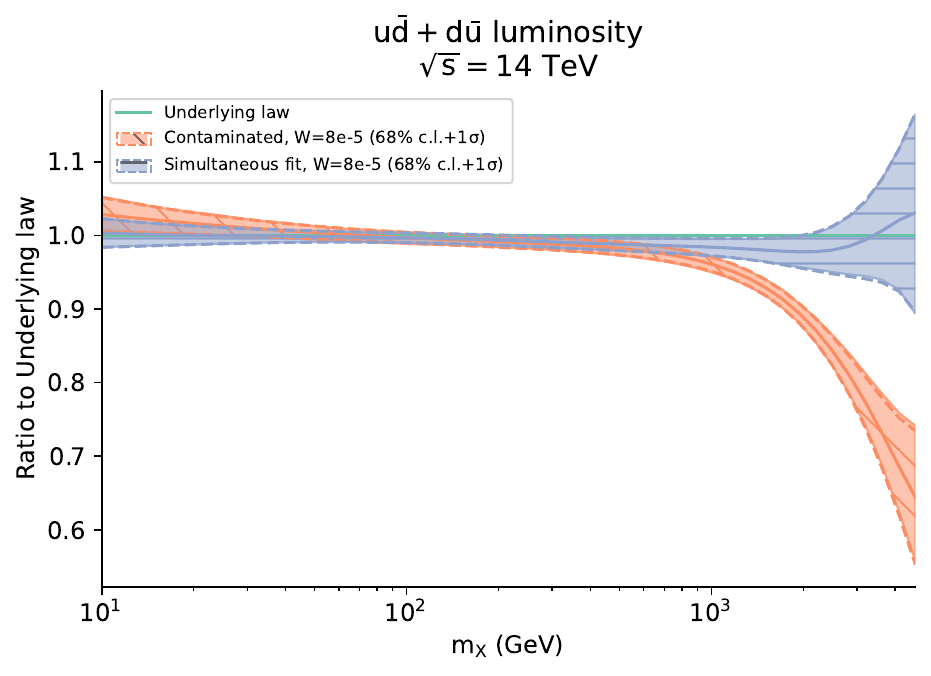}
    \caption{Left: Result of the simultaneous SMEFT fit. Right: Comparison of the simultaneously fitted PDF luminosity (blue) with the contaminated one (orange).}
    \label{fig:simu_fit}
\end{figure}

Another solution is to simultaneously fit the PDFs and the NP degrees of freedom. This can be done using our {\tt SIMUnet} tool~\cite{Costantini:2024xae}. The result of such a fit in our toy model is shown in Fig.~\ref{fig:simu_fit}. We see that the SMEFT fit is able to robustly recover the manually introduced $W'$ contamination. We also observe that the simultaneously fitted PDF (blue) is compatible with the underlying law and seems to be preserved from the contamination effect (orange).

This solution seems quite effective however, it can be difficult to implement. Some non-trivial interplay between PDF and SMEFT needs to be kept under control and a SMEFT sector needs to be identified prior to the fit. Further details will be presented in another future study.

\section*{Acknowledgments}

I am grateful to the organisers of the 58th Rencontres de Moriond for a very well organised conference. I would also like to thank my supervisor Maria Ubiali and the rest of her research group. We are supported by the European Research Council (grant agreement n.950246).

\section*{References}

\end{document}